\documentclass[a4paper,11pt]{article}
\usepackage{pos}
\usepackage{natbib}

\newcommand{\eq}{\begin{equation}}

\newcommand{\feq}{\end{equation}}

\newcommand{\eqn}{\begin{eqnarray}}
\newcommand{\feqn}{\end{eqnarray}}

\newcommand{\be}{\begin{equation}}
\newcommand{\ee}{\end{equation}}
\newcommand{\ben}{\begin{displaymath}}
\newcommand{\een}{\end{displaymath}}
\newcommand{\bea}{\begin{eqnarray}}
\newcommand{\eea}{\end{eqnarray}}

\newcommand{\bean}{\begin{eqnarray*}}
\newcommand{\eean}{\end{eqnarray*}}

\newcommand{\mrm}[1]{\mbox{$\mathrm{#1}$}}

\def\Im{{\mathrm{Im}}}
\def\Re{{\mathrm{Re}}}

\title{Bubbles of nothing and AdS instabilities}

\author*{Nicolò Petri}

\affiliation{Department of Physics, Ben-Gurion University of the Negev, Beer-Sheva 84105, Israel}

\emailAdd{petri@post.bgu.ac.il}

\abstract{In this contribution we review the ideas and results presented in the articles \cite{Dibitetto:2020csn} and \cite{bubbles} where the decay of AdS vacua into bubbles of nothing is studied. We firstly consider AdS vacua in General Relativity with a cosmological constant and in arbitrary dimensions. In this context we study a class of gravitational instantons generalizing the Witten's bubble of nothing and we discuss the derivation of the euclidean action. Secondly we consider a ``stringy'' non-supersymmetric AdS$_4$ vacuum arising in a compactification of massive IIA supergravity on a $S^6$ and preserving G$_2$ symmetry. This vacuum has been shown to have a fully stable KK spectrum and also be protected against brane-jet decays. In this framework we discuss a bubble solution in massive IIA connected to the G$_2$-invariant vacuum and associated to an instability channel. We finally provide the interpretation of this instability in terms of the nucleation of a bubble of nothing dressed up with a D2 charge distribution.}

\FullConference{%
  Corfu Summer Institute 2021 "School and Workshops on Elementary Particle Physics and Gravity"\\
  29 August - 9 October 2021\\
  Corfu, Greece
}

 \tableofcontents

\begin{document}
\maketitle

\section{Introduction}

Providing a microscopic description of classical gravitational systems without supersymmetry is a central challange for the developement of theoretical high-energy physics. Even if it is a well-known fact that supersymmetry constitutes a sufficient condition for non-perturbative stability \cite{Witten:1982df}, the need of a deep understanding of the dynamical supersymmetry breaking mechanism has been manifest since the beginning of research in string theory. This need is of course motivated by the demand of building up phenomenological models that, on one hand, are UV-complete and, on the other hand, can be used to explain experimental evidences. 

Despite the huge effort in the last decades in costructing non-supersymmetric setups in string theory, in the last few years the formulation of Swampland Conjectures radically questioned their reliability as consistent solutions in a quantum theory of gravity (see \cite{Palti:2019pca} for a review of this topic). This approach has been originally motivated by the insights coming from the Weak Gravity Conjecture \cite{ArkaniHamed:2006dz}. Among the various implications of this conjecture one of the sharpest is concerning non-supersymmetric AdS vacua and their expected non-perturbative instability \cite{Freivogel:2016qwc, Ooguri:2016pdq}. This should be realized by a spontanueous nucleation of branes whose emergence completely discharges the flux that supports the vacuum.

Metastable spacetimes and their non-perturbative decays have been object of study for a long time, independently from string theory (see for example \cite{Coleman:1980aw,Brown:1988kg}). The standard approach makes use of the semi-classical approximation to construct gravitational instantons. In this perspective the decay of a spacetime can be viewed as a quantum process of nucleation of a bubble that expands and eventually eats up the original background in a finite time.

Implementing such time-dependent processes in string theory constitutes an extremely complicated challange. The reasons of this are essentialy two. The first is technical since solving the field equations in supergravity without SUSY requires completely different strategies with respect to supersymmetric cases. The second complication is related to the physical interpretation of instantonic geometries in terms of fundamental quantum objects of the string spectrum.
This contribution follows this research trajectory and it is focused on non-perturbative decays of AdS vacua through the nucleation of \emph{bubbles of nothing}. In particular the main purpose of this article is to review the ideas and results of \cite{Dibitetto:2020csn} and \cite{bubbles}.

The decay into bubbles of nothing has been firstly studied in \cite{Witten:1981gj}, where the (in)stability of the KK vacuum in five dimensions was considered. A part this original analysis there are no many examples in the string theory literature of these particular decays. Of particular interest are the two recent works \cite{Ooguri:2017njy} and \cite{GarciaEtxebarria:2020xsr}. In the first one a fully-backreacted solution describing a bubble of nothing in a AdS$_5$ vacuum in M-theory is constructed and, in the second, the authors study these particular instantonic geometries in the context of a Einstein dilaton Gauss-Bonnet model related to heterotic compactifications.

The first article we review in this contribution is focused on decays of AdS spacetimes into bubbles of nothing in the context of General Relativity in arbitrary dimensions and with a cosmological constant \cite{Dibitetto:2020csn}. Even if the results of this paper are not related in principle with any stringy embedding, they allow to identify some crucial properties of these istantonic geometries for AdS vacua. In particular the original idea of \cite{Witten:1981gj} consisting in obtaining the bubble geometry through a double analytic continuation of non-extremal black holes can be applied for a broader class of situations, including a negative cosmological constant and possibly also fluxes. It follows that the presence of a de Sitter foliation characterizing the bubble geometry turns out to be crucial in reproducing the Lorentzian picture of the decay. This is reasonable since the intrinsic time-dependence of de Sitter is the key property allowing to geometrize the surface of an expanding bubble within a higher-dimensional spacetime.

In the second paper reviewed \cite{bubbles} the aforementioned ideas are applied to a non-supersymmetric G$_2$-invariant AdS$_4$ vacuum obtained in a consistent truncation of massive IIA supergravity around a $S^6$ \cite{Lust:2008zd,Cassani:2009ck,Borghese:2012qm,Guarino:2015vca}. This particular AdS$_4$ vacuum raised many interests since it has been shown to be perturbatively stable \cite{Guarino:2020flh} and also protected against brane-jet instabilities \cite{Guarino:2020jwv}. 
This situation is clearly harder to study with respect to the aforementioned case of bubble geometries in GR with a cosmological constant. The equations of motion for the bubble geometry can be solved only with numerical methods and a singular behavior arises requiring an extra effort in interpreting the divergence in terms of brane sources.

A source of inspiration for this anlaysis is the work \cite{Horowitz:2007pr} where the authors study the decay of $\text{AdS}_5\times S^5/\mathbb{Z}_k$ vacua in Type IIB string theory. This non-perturbative process is discussed through various classical regimes that are glued each others by suitable junction conditions. The bubble of nothing geometry turns out to represent an intermediate regime connecting the vacuum with a phase emerging at short distances where the geometry is dominated by a singular behavior associated to istantonic D3 branes.

In \cite{bubbles} an analogous approach is applied to the case of G$_2$-invariant AdS$_4$ vacuum in massive IIA which is also featured by a Freund-Rubin filling flux. The fully-backreacted bubble solution connected to the aforementioned vacuum is worked out numerically and then it is studied by connecting approximated regimes. Crucially a dS$_3$ foliation characterizes the whole radial flow that asymptotically reproduces the vacuum geometry and a singular behavior at short radial distances. This divergent behavior is well-approximated by the geometry of a smeared D2 brane wrapping the dS$_3$.
Crucially an intermediate bubble regime arises interpolating between the vacuum and the D2 source. In this phase the flux contribution is sub-dominant (including Romans' mass) with respect to the contribution coming from the bubble's curvature and a very similar behavior of a bubble of nothing can be recovered. In particular from a 4d perspective the bubble is spontaneously created as a standard bubble of nothing and, corresponding to the initial position of the bubble, a four-dimensional modulus shrinks to zero. Crucially this modulus does not have a geometric interpretation in relation to any cycle within the geometry, but instead to the string coupling.

\section{Bubbles of nothing in gravity theories}\label{cap1}

A predictive framework in studying quantum instabilities of a gravitational system is provided by the semi-classical approximation. In this approach the non-perturbative decay of a gravity vacuum is described in terms of a quantum tunnelling process driven by a gravitational instanton. Intuitively, the instantonic geometry gives the ``classical picture" of a process of nucleation of a bubble that appears spontaneously within an unstable background and expands following a time-dependent dynamics.

A gravitational instanton can be defined as a real and smooth solution of the Euclidean equations of motions approaching the putative vacuum in their asymptotics and characterized by a finite value of the Euclidean action $S_E|_{inst}$. This last property automatically implies that the instanton gives a non-trivial contribution to the path integral in terms of an imaginary part to the energy of the vacuum. This allows to define the decay probability per volume of spacetime as \cite{Coleman:1980aw}
\begin{equation}
 \Gamma/V \sim e^{-\Delta S_E}\,,
\end{equation}
with $\Delta S_E=S_E|_{inst}-S_E|_{vac}$.

In this section we firstly summarize the results of \cite{Witten:1981gj} where the non-perturbative instability of the Kaluza-Klein vacuum in five-dimensional General Relativity is discussed by using the semi-classical approximation. After having introduced and discussed the non-perturbative decay of the KK vacuum into bubbles of nothing, we summarize the main results of \cite{Dibitetto:2020csn} where the aforementioned 5d bubble geometry is extended to General Relativity in arbitrary dimensions and with a negative cosmological constant. In particular we present the analytic form of gravitational instantons for AdS vacua and the change in the euclidean action corresponding to these instantons.

\subsection{The case of the KK vacuum}
\label{bubblenothingwitten}

Wondering about the stability of the KK vacuum $\mathbb{R}^{1,3}\times S^1$ is somehow natural since it is a zero-energy spacetime, but with different asymptotic behaviour of the Minkowski vacuum. For the latter one, the positive energy theorem \cite{Schon:1979rg,Schon:1981vd,Witten:1981mf} ensures the full stability, but the explicit dependence of the energy on the boundary conditions does not allow a direct comparison between the Minkowski and the KK vacua.
In other words we need to take in consideration the KK vacuum by itself and search for istantons describing its decay.
The idea in \cite{Witten:1981gj} is to construct instantons for the KK vacuum by considering non-extremal black holes in 5d and applying a double analytic continuation. The simplest example is provided by the 5d Schwarzschild black hole,
\begin{equation}
 ds^2=-\left(1-\frac{R}{\rho^2}\right)dt^2+\frac{d\rho^2}{\left(1-\frac{R}{\rho^2}\right)}+\rho^2ds^2_{S^3}\,.
\end{equation}
By Wick-rotating the time coordinate  $t\rightarrow i \phi$ it is possible to obtain the Euclidean geometry, 
\begin{equation}
 ds^2_{5,E}=\left(1-\frac{R}{\rho^2}\right)d\phi^2+\frac{d\rho^2}{\left(1-\frac{R}{\rho^2}\right)}+\rho^2ds^2_{S^3}\,.
 \label{eucSchwarzschild}
\end{equation}
The above geometry is well-defined and non-singular in the interval $\rho\in (R, +\infty)$ if and only if the coordinate $\phi$ is periodic as $\phi \sim \phi + 2\pi R$. It is easy to show that for $\rho\rightarrow +\infty$, the solution \eqref{eucSchwarzschild} describes the Euclidean KK vacuum in spherical coordinates, i.e.
\begin{equation}
 ds^2_{\text{5,KKE}}=d\phi^2+d\rho^2+\rho^2ds^2_{S^3}\,,
 \label{euclideanKK}
\end{equation}
where we identified the KK radius $\ell_{\text{KK}}$ with the period of the compact direction of \eqref{eucSchwarzschild}, i.e. $R=\ell_{\text{KK}}$ \cite{Witten:1981gj}. It turns out that the Euclidean solution \eqref{eucSchwarzschild} consitutes an instanton for the KK vacuum, in fact the derivation of the Euclidean action leads to a finite result,
\begin{equation}
 \Delta S_E=\frac{\pi R^2}{4\,G_N^{(4)}}\,.
 \label{bnothingpprob5d}
\end{equation}
We point out that since we are in the semi-classical approximation, this result holds only for values of $R$ that are bigger than the Planck length $\ell_{\text{Pl}}$. This means that the probability of decay is small, but still different from zero.

Let's now study the spacetime in which the KK vacuum decays. To this aim we need to rotate back the Euclidean solution \eqref{eucSchwarzschild} onto the Lorentzian. This time the analytic continuation has to be performed within the $S^3$ in order to keep intact the asymptotic structure of the KK vacuum  \eqref{euclideanKK}.  More concretely, if we parametrize the $S^3$ as $ds^2_{S^3}=d\theta^2+\sin^2\theta\,ds_{S^2}^2$, we can introduce a new time coordinate $\psi$ such that $\theta\rightarrow i\psi+\frac{\pi}{2}$. In this way $ds^2_{S^3}\rightarrow ds^2_{{\scriptsize \mrm{dS}_3}}=-d\psi^2+\cosh^2\psi \,ds^2_{S^2}$ and the Euclidean metric \eqref{eucSchwarzschild} takes the form \cite{Witten:1981gj}
\begin{equation}
 ds^2_{5}=\left(1-\left(\frac{R}{\rho}\right)^2\right)d\phi^2+\frac{d\rho^2}{1-\left(\frac{R}{\rho}\right)^2}+\rho^2ds^2_{{\scriptsize \mrm{dS}_3}}\,.
 \label{bubblenothing5d}
\end{equation}
The background described by \eqref{bubblenothing5d} is non-singular and geodesically complete for $\rho\in(R,+\infty)$.

It is easy to verify that for $\rho\rightarrow +\infty$ we recover the KK vacuum with the coordinates $(\rho,\psi)$ parametrizing a 2d Rindler spacetime \cite{Witten:1981gj}. In other words the coordinates $(\rho,\psi)$ do not cover the whole 2d Minkowski spacetime, but only the exterior of the light cone. This can be easly seen since, in the limit $\rho\rightarrow +\infty$, the asymptotic background is defined for $\rho\in (0,+\infty)$ and this implies that $(\rho,\psi)$ cover only the region $x^2-t^2=\rho^2>0$ with $x=\rho\cosh\psi$ and $t=\rho\sinh\psi$. Moving away from the asymptotics, the 2d Rindler spacetime is distorted by the warp factors and the coordinate $\rho$ is now defined on the interval $(R, +\infty)$. The coordinates $(\rho,\psi)$ now parametrize the region $x^2-t^2>R^2$ on which the 5d background \eqref{bubblenothing5d} turns out to be non-singular and geodesically complete. Crucially this is due to the presence of compact KK coordinate $\phi$ that shrinks to zero size at $\rho=R$ in such a way the boundary defined by the hyperbola of radius $R$ is smoothly sealed off \cite{Witten:1981gj}.

We can then interpret the spacetime \eqref{bubblenothing5d} as a bubble expanding in the time $\psi$ whose geometry is described by a 3d de Sitter manifold $\mrm{dS}_3$. The bubble has only an exterior since by construction the metric \eqref{bubblenothing5d} describes a non-singular geodesically complete spacetime for values of $(\rho,\psi)$ covering only the external region of the hyperboloid $x^2-t^2=R^2$. For this reason, this quantum instability of the 5d KK vacuum has been called {\itshape bubble of nothing} in \cite{Witten:1981gj}.

Since any decay process has to preserve energy, it follows that the spacetime \eqref{bubblenothing5d} has zero energy and this means that a positive energy theorem for the KK vacuum does not exist: there exists a solution with zero energy asymptotically approaching the KK vacuum and this is exactly what allows the KK vacuum to decay \cite{Witten:1981gj}. This fact can be also seen by looking at the surface integral defining the energy. In particular, taking the zero-time surface at $\psi=0$, the asymptotics \eqref{bubblenothing5d} gives contributions to the KK vacuum of the order $\rho^{-2}$ and it can be shown that, for a 4d observer, only terms of the order $\rho^{-1}$ give positive contributions to the energy of the vacuum \cite{Witten:1981gj}. 

Finally we point out that in the decay a change of topology occurs since at $\psi=0$ the bubble of nothing is described by the topological space $\mathbb{R}^2\times S^2$. The change of topology is intimately related to the fact that the instability has been obtained by a double analitical continuation of a non-extremal black hole and it has very interesting consequences when one tries to introduce spinors \cite{Witten:1981gj}. In fact the space $\mathbb{R}^2\times S^2$ is simply-connected and this implies that there exists a unique spin structure defined on it. On the other hand the KK vacuum at $t=0$, $\mathbb{R}^3\times S^1$, is not simply-connected and this implies that the spinors are defined up to a phase $\alpha$ as it follows \cite{Witten:1981gj}
\begin{equation}
 \psi(x,\phi)=\sum_n\,\psi_n(x)\,e^{\frac{i}{R} \left(n-\frac{\alpha}{2\pi}\right)\phi}\,.
\end{equation}
It easy to see that the existence of covariantly constant spinors on $\mathbb{R}^3\times S^1$ imposes $\alpha=0$ while
the existence of the spin structure on the $S^2$ requires that $\alpha=\pi$, namely anti-periodic boundary conditions \cite{Witten:1981gj}. This consideration implies that the gravitational instanton \eqref{eucSchwarzschild} does not contribute to the path integral of theories described by covariantly constant spinors. This fact is quite interesting since it tells us that, with suitable boundary conditions on fermions, we can ``cure" the quantum instability of the KK vacuum and, moreover, it tells us that these boundary conditions correspond exactly to the requirement of the existence of covariantly constant spinors \cite{Witten:1981gj}.

\subsection{Bubbles of nothing within AdS vacua}

It is natural to wonder if the idea and results of \cite{Witten:1981gj} can be implemented in the case of vacua with constant curvature in $(D+1)$ dimensions \cite{Godazgar:2009fi}. Of course the main difference with respect to section \ref{bubblenothingwitten} is the inclusion of a negative cosmological constant\footnote{A discussion on the case of positive cosmological constant can be found in \cite{Dibitetto:2020csn}.} $\Lambda$. In addition to this we also allow the asymptotic region to be described not only by a single sphere but rather a product of them. 

Let's start by discussing the vacuum geometries.
The zero-curvature case is described by a Ricci-flat vacuum solution with zero-energy and extends the 5-dimensional case presented in \cite{Witten:1981gj} to $D+1$ dimensions. When $\Lambda$ is running we will obtain bubble geometries that reproduce asymptotically locally AdS$_{D+1}$ vacua written in the form
\begin{equation}
 \begin{split}
ds^2_{D+1}=&\left(1-k \rho^2\right)d\phi^2 + \left({D-2\over d-1}\right) {d\rho^2 \over \left(1-k \rho^2\right)} + \rho^2\left(L^2_{{\rm dS}_d}ds^2_{{\rm dS}_d} + L^2_{S^{D-d-1}}ds^2_{S^{D-d-1}}\right)\,, \\
k=& {2(D-2) \over D(D-1)(d-1)}\Lambda\,,
\end{split}
\label{falseDplus1}
\end{equation}
where $\phi$ parametrizes the KK circle\footnote{Even if this $S^1$ is not technically compact in the case of AdS, we will still refer to it as a KK circle.} with period $2\pi \ell_{\text{KK}}$. The elements $ds^2_{{\rm dS}_d}$ and $ds^2_{S^{D-d-1}}$ describe respectively a $d$-dimensional de Sitter geometry and a ($D-d-1$)-dimensional sphere with unit radius.  The cosmological constant can be expressed in terms of the Ricci scalar as it follows,
\begin{equation}
|\Lambda| = {D(D-1) \over 2 \ell^2}, \qquad R={2(D+1) \over D-1} \Lambda\,,
\end{equation}
with $\ell$ radius of AdS$_{D+1}$.
The above solutions represent non-singular and geodesically complete solutions of Einstein equations $R_{\mu\nu}={2 \over D-1}\Lambda g_{\mu\nu}$ when the conditions
\begin{equation}
L_{{\rm dS}_d}=1, \qquad  L_{S^{D-d-1}}=L_{{\rm dS}_d}\sqrt{{D-d-2 \over d-1}}\,.
\label{radiiconstraint}
\end{equation}
hold\footnote{We point out that we need $D\neq d+2$ and $d\ge2$ for the $(D+1)$-dimensional solution to exist.}. 
We point out that the case $D=d+1$ is particularly interesting. This is the situation in which the foliation with $S^{D-d-1}$ is not present. In this case the geometry \eqref{falseDplus1} reproduces the topology of AdS. When the additional $S^{D-d-1}$ is present the geometry is only locally $\mathrm{AdS}_{D+1}$. 

The peculiar foliation of \eqref{falseDplus1} in terms of dS slicings is crucial in order to reproduce istantonic bubble geometries. We can continue the geometry \eqref{falseDplus1} to a Euclidean metric and think in analogy to the 5d case of section \ref{bubblenothingwitten}. It is thus possible to show that the asymptotic behavior of \eqref{falseDplus1} can be reproduced as the $\rho \to \infty$ limit of a Euclidean black hole featured by a product of two spheres at the horizon \cite{Dibitetto:2020csn},
\begin{equation}
\begin{split}
 ds^2_{E,D+1}=& f(\rho)d\phi^2+\left(\frac{D-2}{d-1}  \right)\frac{d\rho^2}{f(\rho)} +\rho^2\left(ds^2_{S^d} + \left({D-d-2 \over d-1}\right) ds^2_{S^{D-d-1}}\right)\,, \\
f(\rho)=&1 -k \rho^2 - \left({R\over \rho}\right)^{D-2}\,.
\end{split}
 \label{bubblenothingeuclid}
\end{equation}
In the corresponding Lorentzian black hole solution, there is a coordinate singularity at $\rho=\rho_0$, where $f(\rho_0)=0$. In order to avoid conical singularities, we need to impose the condition
\begin{equation}
\ell_{\text{KK}}^2 = {4(D-2) \over f'(\rho_0)^2(d-1)}\,.
\label{lkktorho0cond}
\end{equation}
This condition leaves the geometry \eqref{bubblenothingeuclid} well-defined and non-singular for $\rho \in (\rho_0, +\infty)$. 

We point out that that continuing the sphere $S^d$ to dS$_d$ produces an instanton geometry describing an expanding bubble whose surface is given by $\mrm{dS}_d\times S^{D-d-1}$ \cite{Dibitetto:2020csn},
\begin{equation}
ds^2_{D+1}=f(\rho)d\phi^2+\left(\frac{D-2}{d-1}  \right)\frac{d\rho^2}{f(\rho)} +\rho^2\left(ds_{{\rm dS}_d}^2 + \left({D-d-2 \over d-1}\right) ds^2_{S^{D-d-1}}\right)\,.
 \label{bubblenothingKK}
\end{equation}
Asymptotically the geometry \eqref{bubblenothingKK} reproduces the vacuum \eqref{falseDplus1} and the full spacetime background is defined for $\rho \in (\rho_0, + \infty)$.  Thus we obtained an extension of the bubble of nothing that describes the decay of the vacua \eqref{falseDplus1}.
We finally point out that in the case of zero-curvature the relation \eqref{lkktorho0cond} reduces to \cite{Dibitetto:2020csn}
\begin{equation}
 \ell_{\text{KK}}^2 = {4 R^2 \over (d-1)(D-2)}\,,\label{lKKzerocurvature}
\end{equation}
where $\rho_0=R$.
The 5d case of \cite{Witten:1981gj} can be thus recovered by requirng that $D=4\,,\,\, d=3$, namely $R=\ell_{\text{KK}}\,$.

\subsection{Euclidean action}

In this section we derive the change in the Euclidean action for the bubble geometry \eqref{bubblenothingKK}. It is well known that in order to reproduce Einstein equations we need to include the Gibbons-Hawking-York (GHY) term describing the boundary contributions that are not fixed by the requirement of vanishing variations at the boundary. In our particular situation the boundary at infinity and the instanton geometry of the bubble constitute boundaries where we need to evaluate the GHY action.

The Euclidean action has the following form
\begin{equation}
 \begin{split}
  &S_E=S_{E, \,\text{bulk}}+S_{E,\, \text{GHY}}\,,\\
  &S_{E,\, \text{bulk}}=\frac{1}{2\kappa_{D+1}^2}\int{d^{D+1} x_E\,\sqrt{g_{D+1}}\,\left (R-2\Lambda \right)}\,,\\
  &S_{E, \,\text{GHY}}=\frac{1}{\kappa_{D+1}^2}\int{d^{D} y_E\,\sqrt{h_{D}}\,\theta_{D}}\,,
 \end{split}\label{euclidactionKK}
\end{equation}
where the coordinates $\{y\}$ parametrize any $D$-dimensional boundary of the background, $h_{D}$ is the induced metric and $\theta_{D}$ is the trace of the extrinsic curvature of the boundary. 
Let's consider the cases of zero and non-zero curvature separately\footnote{For a detailed derivation of $\Delta S_E$ see section 3.1 in \cite{Dibitetto:2020csn}.}.
\paragraph{Zero curvature case: $\Lambda = 0$.}  In this case $\Delta S_{E, \,\text{bulk}}$ vanishes, since it turns out to be proportional to $\Lambda$, moreover  $\rho_0 =R$. The bounce takes the form \cite{Dibitetto:2020csn}
\begin{equation}
 \Delta S_E^{(\Lambda=0)} =   {\pi \ell_{\text{KK}} \over \kappa_{D+1}^2}\sqrt{{d-1\over D-2}}\left({D-d-2 \over d-1}\right)^{(D-d-1)/2} (D-2)R^{D-2}\, {\rm vol}_{S^d} {\rm vol}_{S^{D-d-1}} \,.
\end{equation}
In the case $D=4$ and $d=3$, the $S^{(D-d-1)}$ disappears and the expression boils down to the 5d result \eqref{bnothingpprob5d}. Thi can be seen by using the standard relation defining the Planck masses under dimensional reduction
\begin{equation}
{1\over \kappa_D^2} = {1 \over 8 \pi G_D} = {2\pi \ell_{\text{KK}} \over \kappa_{D+1}^2} \, .
\end{equation} 
It is interesting to look at the length scales contributing to this decay rate.  We can write \cite{Dibitetto:2020csn},
\begin{equation}
\Delta S_E^{(\Lambda=0)} \sim {\ell_{\text{KK}} \over \ell_{\text{Pl}}} \left({R \over \ell_{\text{Pl}}}\right)^{D-2} \sim \left({R \over \ell_{\text{Pl}}}\right)^{D-1}\, ,
\end{equation}
where we used \eqref{lKKzerocurvature} and $\rho_0=R$. We thus have a positive bounce that is large whenever the semiclassical approximation holds and that gives rise to a decay rate exponentially suppressed.
\paragraph{Non-zero curvature case: $\Lambda < 0$.} In this case we have non-zero contributions both from the boundary and from the bulk. One obtains the following result \cite{Dibitetto:2020csn},
\begin{equation}
\begin{split}
\Delta S_E^{(\Lambda\neq0)} =&  \,{\pi \ell_{\text{KK}} \over \kappa_{D+1}^2} {\rm vol}_{S^d} {\rm vol}_{S^{D-d-1}}\sqrt{{d-1\over D-2}}\left({D-d-2 \over d-1}\right)^{(D-d-1)/2} \\
&\times \Big[(D-2)R^{D-2} - 4k\rho_0^D\Big] \,,
\end{split}
\end{equation}
where it has been used that $f(\rho_0)=0$. We point out that the second term in the above expression represents the new contribution proportional to $\Lambda$.
Looking at the length scales we have that \cite{Dibitetto:2020csn}
\begin{equation} 
\Delta S_{E}^{(\Lambda\neq 0)} \sim {\ell_{\text{KK}} \over \ell_{\text{Pl}}} \left( \left({R \over \ell_{\text{Pl}}}\right)^{D-2} + {4\over D-2} \left({\rho_0 \over \ell_{\text{Pl}}}\right)^{D-2} \left({ \rho_0 \over \ell}\right)^2 \right) \, .
\end{equation}
We conclude that the decay rate is exponentially suppressed when the semiclassical approximation is valid.

\section{Bubble instabilities in massive IIA vacua}

In this section we approach the study of non-perturbative instabilities of AdS vacua in string theory by summarizing the main results of \cite{bubbles}. In this work the non-perturbative (in)stability of a non-supersymmetric AdS$_4\times S^6$ vacuum with residual G$_2$-symmetry is studied in the low-energy framework of massive IIA supergravity.

The main result of \cite{bubbles} consists in the numerical derivation of the fully-backreacted solution in 10d describing the instability of the aforementioned AdS$_4$ vacuum. This supergravity solution is then studied in terms of various regimes approximating the phases of the decay process. This approach takes direct inspiration from the seminal work \cite{Horowitz:2007pr} where the authors study the instabilities of AdS$_5\times S^5/\mathbb{Z}_k$ vacua of Type IIB through different approximated regimes associated to the bounce geometry. In particular, in \cite{Horowitz:2007pr} the authors were able to individuate a discharged regime described by the typical behavior of an expanding bubble of nothing and a singular regime where the bubble geometry is modified by the contributions of the $F_5$ flux and a euclidean D3 brane source is driving the decay.

In \cite{bubbles} a similar analysis allows to individuate a particular regime in which the complete bounce solution is consistently approximated by an expanding \emph{dilaton bubble} within the AdS$_4\times S^6$ vacuum. Then another regime emerges at short distances where a D2 singularity dominates the flow.
In this section we introduce firstly the concept of dilaton bubble \cite{bubbles} as the instanton geometry defining the bubble regime for the G$_2$-invariant AdS$_4$ vacuum. We thus formulate the 10d Ansatz and we present the numerical solution of massive IIA supergravity describing the instability of the aforementioned vacuum. Finally we discuss the various regimes of the decay with particular attention to the intermediate bubble regime.

\subsection{The dilaton bubble} \label{dilatonbubble}

Let's consider an istantonic geometry reproducing an approximated regime in which the fluxes and Romans' mass give a negligible contribution to the stress-energy tensor. To this aim we can use M-theory as a seed framework where to construct the bubble geometry \cite{bubbles}.
In particular we can consider the embedding of a 5d Schwarzschild black hole in eleven dimensional supergravity by taking a Ricci-flat manifold $M_6$ as the internal space. The metric is of the form,
\begin{equation}
d s_{11}^2 \, = \, -\left(1-\left(\frac{R}{\rho}\right)^2\right)d\tau^2 \, + \, \frac{d\rho^2}{1-\left(\frac{R}{\rho}\right)^2} \, +\, \rho^2d s_{S^3}^2 \, + \,  d s_{M_6}^2 \ ,
\end{equation}
where $R$ is the Schwarzschild radius. As we discussed in section \ref{cap1} we can obtain a bubble geometry by performing the double analytic continuation $\tau \ \rightarrow \ i\,\psi$ and $S^3 \ \rightarrow \ \mathrm{dS}_3$
where $\psi$ is periodic. If one reduces on $S^1_{\psi}$, the 10d background takes the form \cite{bubbles}
\begin{equation}\label{BoN_Schw_gauge}
	\begin{aligned}
			d s_{10}^2 & =  H^{-1/2}d \rho^2 \, +\, \rho^2H^{1/2}d s_{\mathrm{dS}_3}^2 \, + \, H^{1/2}d s_{M_6}^2 \ , \\
		e^{\Phi} & =  H^{3/4}\,,
	\end{aligned}
\end{equation}
\begin{figure}[!htb]
	\centering
	\includegraphics[width=0.6\textwidth]{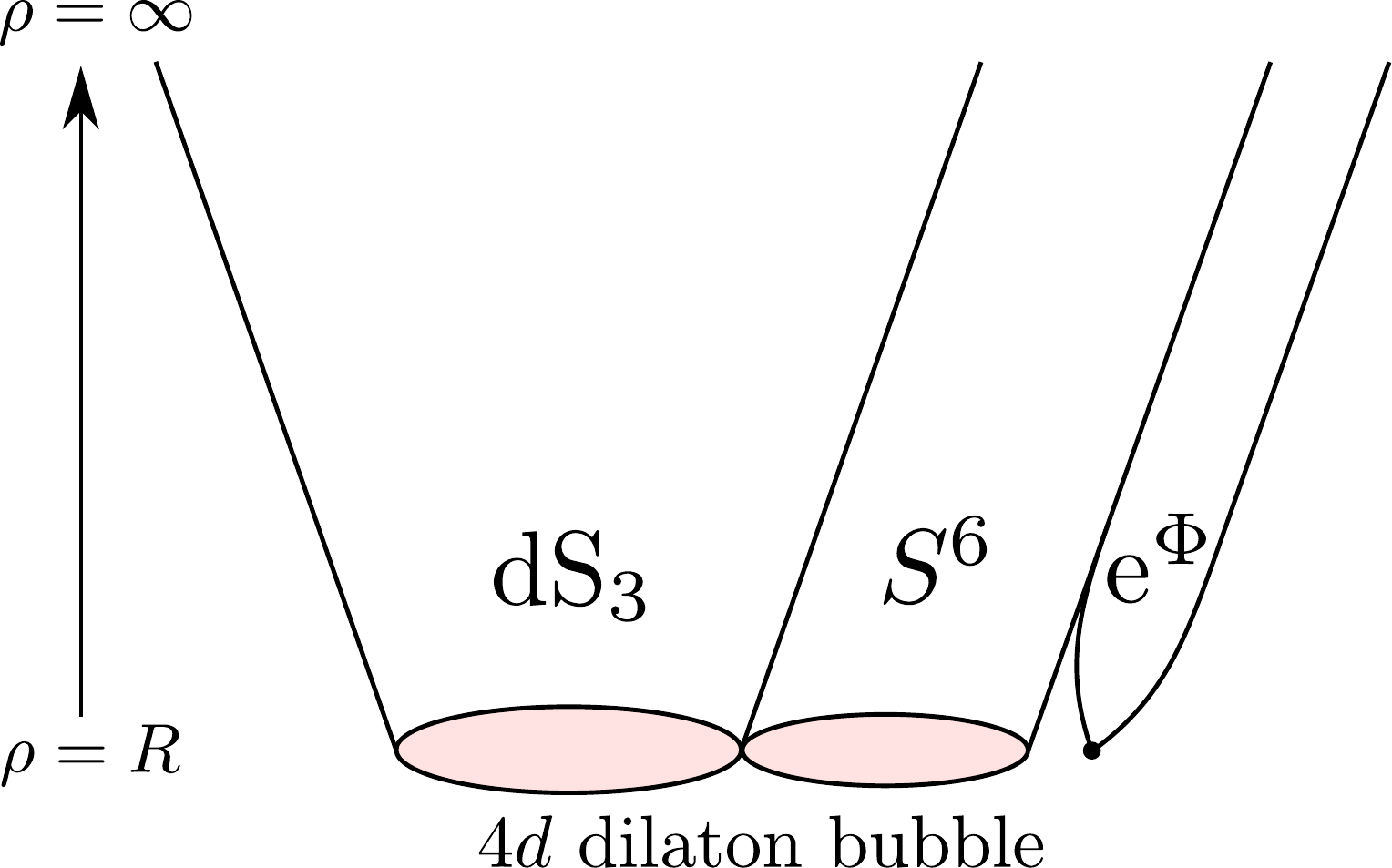}
	\caption{\it The artistic impression of the "dilaton bubble" constructed in \cite{bubbles}.}
	\label{fig:Dilaton_bubble}
\end{figure} 
with $H=1-\left(\frac{R}{\rho}\right)^2$ and $\rho\,\in\,[R,+\infty)$. From the 10d point of view, this geometry has the form of a bubble of nothing where instead of a physical cycle within the geometry, it is the string coupling $e^{\Phi}$ that shrinks. In \cite{bubbles} the uniqueness of such a bubble geometry is also discussed and it is shown that \eqref{BoN_Schw_gauge} represents the unique background admitting a frame where both dS$_3$ and $M_6$ have a finite size at the surface of the bubble $\rho=R$.
This frame is represented by the 11d Einstein frame. Taking the perspective of an observer living within the four-dimensional vacuum, it experiences a bubble of nothing geometry with radius $R$ which is created at the center of the space and immediately starts to expand, eventually eating up the entire vacuum (see figure \ref{fig:Dilaton_bubble}).
\begin{figure}[!htb]
	\centering
	\includegraphics[width=0.7\textwidth]{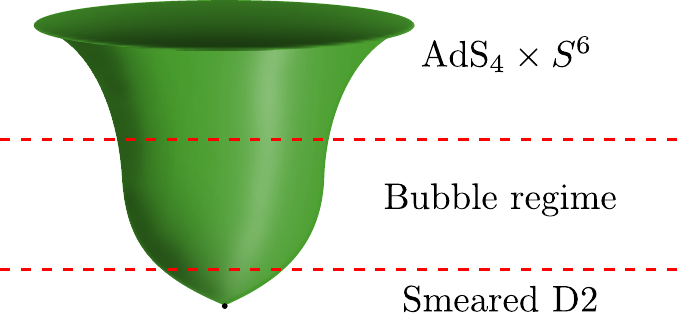}
	\caption{\it The picture of the radial evolution of the bubble geometry taken from \cite{bubbles}. The "dilaton bubble" is a only an intermediate description. In the asymptotics it reproduces the vacuum geometry. At small distances it ultimately flows to a smeared D2 brane source.}
	\label{fig:Dbubble}
\end{figure} 
From the 10d point of view, the bubble geometry has a different origin than the standard bubble of nothing, since the string coupling rather than an internal cycle shrinks.

As we mentioned at the beginning of this section, our aim is the study of non-perturbative instability of certain AdS$_4$ vacua obtained from compactifying massive IIA supergravity on a $S^6$. As we are going to show the aforementioned dilaton bubble solution is a fundamental building block in the decay process. Such a bubble geometry constitutes only an approximate description in an intermediate regime when the full geometry of the decay is considered in massive IIA supergravity. 
This is due to the presence of extra ingredients that crucially imply new emergent features. In particular, the presence of AdS$_4\times S^6$ vacuum requires a different behavior at large distances and a non-zero Freund-Rubin flux implies a modified short distance behavior induced by the presence of smeared D2 brane sources. Furthermore, the non-zero Romans' mass and the departure from Ricci-flatness of $M_6$ determine two other complications. Even if these two elements are both fundamental for the existence vacuum in the asymptotics, in the intermediate phase these ingredients only yield subleading modifications of the bubble geometry. A representation of the different regimes of the complete geometry of the instability is given in figure \ref{fig:Dbubble}.

\subsection{Ten-dimensional Ansatz and G$_2$-invariant vacuum}

We start our analysis by formulating an Ansatz for solutions in massive IIA describing the bounce geometry for AdS$_4\times S^6$ vacua\footnote{The analysis can be extended to six-dimensional compact nearly-K\"ahler manifolds \cite{bubbles}.}. Taking the ispiration from \cite{Witten:1981gj} and from our discussion in section \ref{cap1}, we search for bounce geometries whose 4d external part is represented by a domain wall with a worldvolume curved by a dS$_3$ geometry. In particular we are interested in the situation where the domain wall geometry preserves the G$_2$ isometries that act on the 6-sphere as $S^6\simeq {\mathrm G}_2/{\mathrm SU}(3)$.
The most general 10d background of this type has the form \cite{bubbles}
\begin{equation}\label{10ansatz} 
\begin{split}
	&ds_{10}^2 \,=\, e^{2V(r)}\left( d r^2 + L^2e^{2A(r)} d s^2_{\text{dS}_3}  + g^{-2}ds_{S^6}^2 \right)\,,\\[1mm]
	&e^{\Phi} \,=\, e^{\phi(r)}\,,\qquad \qquad H_3 = d B_2 =   g^{-2}\,b'\,d r\wedge J+ 3\, g^{-2} b \,\Re\,\Omega\,,\\
	&F_0 \,=\, m\,,\qquad \qquad F_2 \,= \, F_0 B_2\,,\\ 
	&F_4 \,=\, f_6(r) \star \text{vol}_{S_6} + f_{41}(r)J\wedge J + f_{42}(r)\,d r\wedge \Im\Omega\,,
	\end{split}
\end{equation}
where $m$ is the Romans' mass, $g$ is an overall scale for the internal manifold controlling the $F_6$ flux and $L$ turns out to be related to the radius of AdS$_4$ asymptotic vacuum. The prime denotes the derivative with respect to the coordinate $r$.

The real 2-form $J$ and the complex 3-form $\Omega$ define the nearly-K\"ahler structure on $S^6$ (for more details see appendix B of \cite{bubbles}). In particular the NSNS field strength is defined by the B-field $B_2 =g^{-2}\, b(r) J$ and the RR fluxes can be expressed in terms of the functions $b(r)$ and $\zeta(r)$ as it follows
\begin{equation}\label{bianchifluxes}
\begin{split}
&f_6 = -\frac{1}{g^2}\big( m b^3 + 6g b \zeta + 5g \big)\,,\quad f_{41} = \frac{1}{2g^4} \big( m b^2 + 2g\zeta \big)\,,\quad f_{42} = \frac{1}{2g^3} \,\zeta^\prime\,.
\end{split}
\end{equation}
We point out that the pure gauge function in front of the $dr^2$ term in the metric has been chosen in such a way to describe the space transverse of D2 branes. The reason of this is due to the Freund-Rubin term $F_6=-5 g^{-1}\,\text{vol}_{S^6}+\cdots$ hinting a special role of these sources in the backgrounds captured by the Ansatz \eqref{10ansatz}.

The above 10d Ansatz is included in a consistent truncation of massive IIA supergravity on nearly-K\"ahler manifolds \cite{Kashani-Poor:2007nby,Cassani:2009ck}. From a 4d perspective the physics of \eqref{10ansatz} is described by five real scalar fields\footnote{The scalar field $\xi$ constitutes a flat direction for the 4d scalar potential \cite{Kashani-Poor:2007nby,Cassani:2009ck}.} $(U,\phi,b,\xi,\zeta)$ with $U=V-\frac{\phi}{4}$. The geometry takes the form of a curved domain wall with the form \cite{bubbles}
\begin{equation}
 ds^2_4=e^{8U(r)}\,\left(dr^2+L^2 e^{2A(r)} \right)\,.
\end{equation}
This 4d model contains three different AdS$_4$ vacua \cite{Cassani:2009ck} that can be realized locally by chosing
\begin{equation}
 e^A=e^{-4U_0}\,\sinh\left(\frac{r}{e^{-4U_0}L}\right)\,.
\end{equation}
We note that the particular foliation of AdS$_4$ we are considering, namely by using dS$_3$ slicing of the 4d spacetime, is obviously crucial in order to study the bounce geometries within the aforementioned vacua. As we said we are interested in a particular realization of AdS$_4$ without supersymmetry and preserving the residual symmetry $G_2$, namely\footnote{For simplicity of notation we fixed $m=1$ and $g=\frac{1}{2}$.} \cite{Lust:2008zd,Cassani:2009ck}
\begin{equation}
 e^{2\phi}=\frac{\sqrt 3}{2}\,,\qquad e^{8U_0}=\frac{3\sqrt 3}{8}\,,\qquad \zeta=-1\,,\qquad b=-1\,,\qquad L^2=\frac{3\sqrt 3}{4}\,.
\end{equation}
This vacuum has been shown to be perturbatively stable against all the scalar fluctuations of 4d maximal ISO$(7)$ supergravity \cite{Guarino:2015vca} and also including the full KK spectrum of massive IIA supergravity \cite{Guarino:2020flh}. Moreover it has been recently shown that this vacuum is stable against brane-jet instabilities \cite{Guarino:2020jwv}. For these reasons this particular AdS$_4$ vacuum is of particular interest for our purposes. In fact it constitutes the ideal stringy setup where we can try to apply the ideas discussed in section \ref{cap1} and try to study its non-perturbative (in)stability.

\subsection{The decay channel}

In order to search for the bounce solution describing the instability of the G$_2$-invariant vacuum we are forced to use numerical methods. From this perspective the boundary behavior at the end of the radial flow turns out to be crucial to understand the physical features of the decay. We firstly point out that the presence of the Freund-Rubin flux filling the $S^6$ induces singular terms into the equations of motion in the limit where the internal space shrinks. This means that we must include a source at small distances. This may also be argued from the fact that the absence of a source would imply a vanishing $F_6$ through the 6-sphere, trivializing our setup. From this it follows that the inclusion of D2 sources filling dS$_3$ would restore charge conservation in the system.

\begin{figure}[!http]
	\centering
	\includegraphics[width=0.98\textwidth]{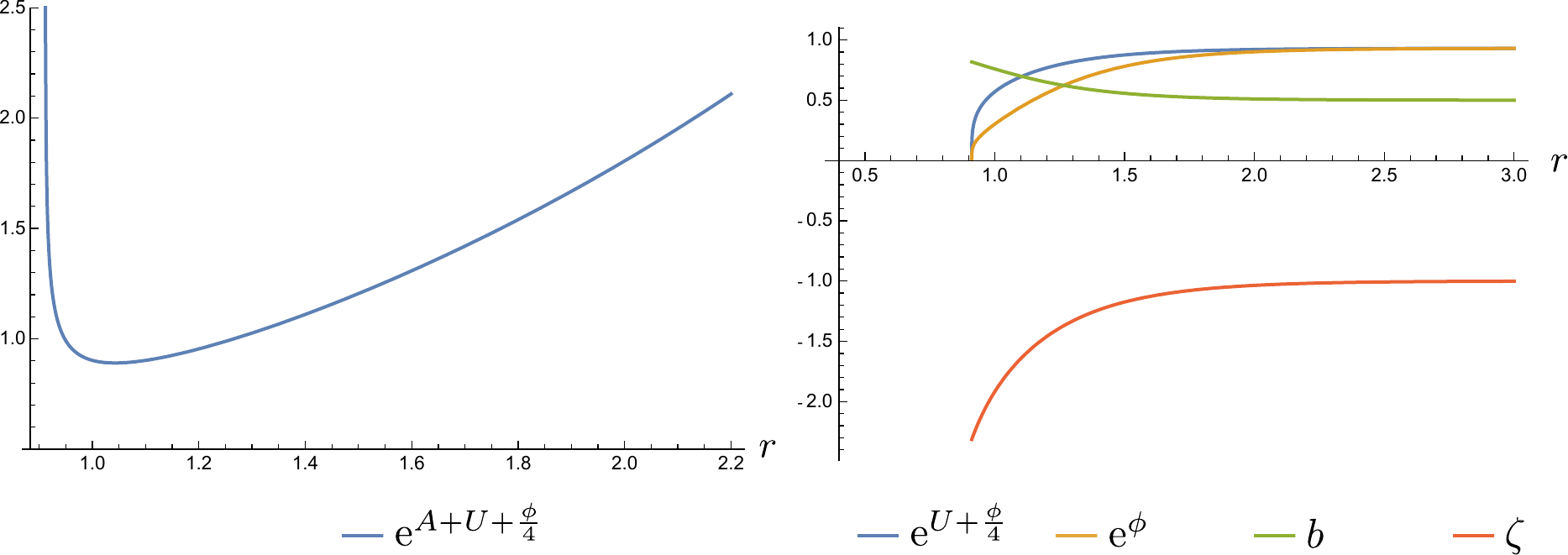}
	\caption{\it The radial solution describing the instability of the non-SUSY G$_2$-invariant vacuum. The factors $e^{A+U+\frac{\phi}{4}}$ and $e^{U+\frac{\phi}{4}}$ respectively describe the size of dS$_3$ and $S^6$. Plot taken from \cite{bubbles}.}
	\label{fig:DbubblePlot}
\end{figure}

We can obtain the initial condition for our numerical integration by considering the fluctuations around the AdS$_4$ vacuum by writing the metric in the usual Fefferman-Graham form and demanding that only normalizable modes are activated. If one diagonalizes the mass matrix for the G$_2$-invariant vacuum, the eigenvalues are given by
\begin{equation}
 M^2 L^2=6,\,6,\,20,\,20\,.
\end{equation}
From the the fundamental relation $\Delta (\Delta-3)=M^2\,L^2$ we can obtain the corresponding conformal dimensions,
\begin{equation}
 \Delta= \frac{3+\sqrt{33}}{2},\qquad\frac{3+\sqrt{89}}{2}\,,
\end{equation}
each of them with multiplicity 2, namely associated to two-dimensional eigenspaces. We can thus derive the asymptotic behaviors of our fields $U, b, \phi, \zeta$ \cite{bubbles}, 
\begin{equation}\label{asymptotic_expr_scalars}
\begin{split}
&U \,=\, U_0 \,+\,  \frac{4c_1-3c_2}{28}\, z^{\frac{3+\sqrt{33}}{2}} + c_4\,z^{\frac{3+\sqrt{89}}{2}} + \ldots \,,\\
&b \,=\, b_0 \,+\,  \frac{3 c_2-4c_1}{7}\, z^{\frac{3+\sqrt{33}}{2}} + \frac{7c_3 + 36 c_4}{12}\,z^{\frac{3+\sqrt{89}}{2}} + \ldots\,,\\
&\phi \,=\, \phi_0 \,+\, c_2\, z^{\frac{3+\sqrt{33}}{2}} - c_3\,z^{\frac{3+\sqrt{89}}{2}} + \ldots \,,\\
&\zeta\,=\, \zeta_0 \,+\,  c_1\, z^{\frac{3+\sqrt{33}}{2}} + c_3\,z^{\frac{3+\sqrt{89}}{2}} + \ldots\,,
\end{split}
\end{equation}
where $z=e^{-r/\ell}$ with $\ell=e^{-4U_0}L=\sqrt{2}$ in our conventions. We use the above expression as initial conditions on the fields for the numerical integration. By construction the four coefficients $c_i$ activate only the normalizable modes and, for any choice of them, one can observe that the flow always ends up in a smeared D2 brane singularity.
Among various choices for the coefficients $c_i$, one can impose some particular constraints in order to reproduce the bubble geometry \eqref{BoN_Schw_gauge} as an intermediate regime of our numerical solution. In particular, let's focus on the choice
\begin{equation}
 3U-\frac{\phi}{4}=\text{constant}\,,
\end{equation}
implying $3c_1-4c_2=0$ and $c_3+12c_4=0$. This choice leaves only two independent parameters that turn out to fix the position of the D2 source $r_{\text{D}2}$ and the radius of the bubble $R$. The result of this integration is given in figure \ref{fig:DbubblePlot}. 

This numerical result can be understood by introducing three different regimes: the asymptotic AdS$_4$ vacuum, a source regime at low distances and an intermediate bubble regime. Starting by perturbing the AdS vacuum with normalizable modes, the solution depicted in figure \ref{fig:DbubblePlot} exhibits a regime that can be approximated by the dilaton bubble geometry \eqref{BoN_Schw_gauge}. Such a solution exists only as an approximated solution in the regime where one can neglet the contributions of fluxes as well as the curvature of the internal $S^6$. As we discussed in section \ref{dilatonbubble} the crucial property of this bubble geometry is determined by the dilaton that shrinks smoothly at the location of the bubble.

The presence of Romans' mass and fluxes implies that the aforementioned bubble regime must be modified along the flow. In the case of geometry \eqref{BoN_Schw_gauge} the object that must be included at short distances is an euclidean D2 brane smeared along the $S^6$. The geometry describing this third phase has the following form
\begin{equation}\label{smearedD2}
\begin{split}
&d s_{10}^2  = H_{\mathrm{D}2}^{-1/2}d s_{\mathrm{dS}_3}^2\,+\, H_{\mathrm{D}2}^{1/2}\left(d r^2\,+\, g^{-2}d s_{S^6}^2\right) \ ,\\
&e^{\Phi}  = H_{\mathrm{D}2}^{1/4} \ ,\qquad C_{3}  = \left(H_{\mathrm{D}2}^{-1}-1\right) \, \mathrm{vol}_{\mathrm{dS}_3}\ ,\\\vspace{0.5cm}
&\text{with}\qquad H_{\mathrm{D}2} = Q_{\mathrm{D}2}(r-r_{\mathrm{D}2})\,,
\end{split}
\end{equation}
where $Q_{\mathrm{D}2}=5 g$ is fixed by the Page $F_6$ at infinity and $r_{\text{D}2}$ is fixed by the dynamics of the solution. The expressions \eqref{smearedD2} constitute an approximated solution like the intermediate bubble regime and represent a good approximation of our numerical flow in the limit where the curvatures of dS$_3$ and $S^6$ are negligible.

The three regimes can be glued together consistently by rewriting the bubble regime \eqref{BoN_Schw_gauge} in a gauge compatible with the D2 geometry \eqref{smearedD2}. This can be achieved introducing the new coordinate
$\rho^2=R^2+(r-r_{\text{B}})^2$ with $r_{\text{B}}$ integration constant associated to the position of the bubble wall\footnote{For a more detail analysis on the gluing between the different regimes we refer to \cite{bubbles}}. 

\begin{figure}[!http]
	\centering
	\includegraphics[width=0.75\textwidth]{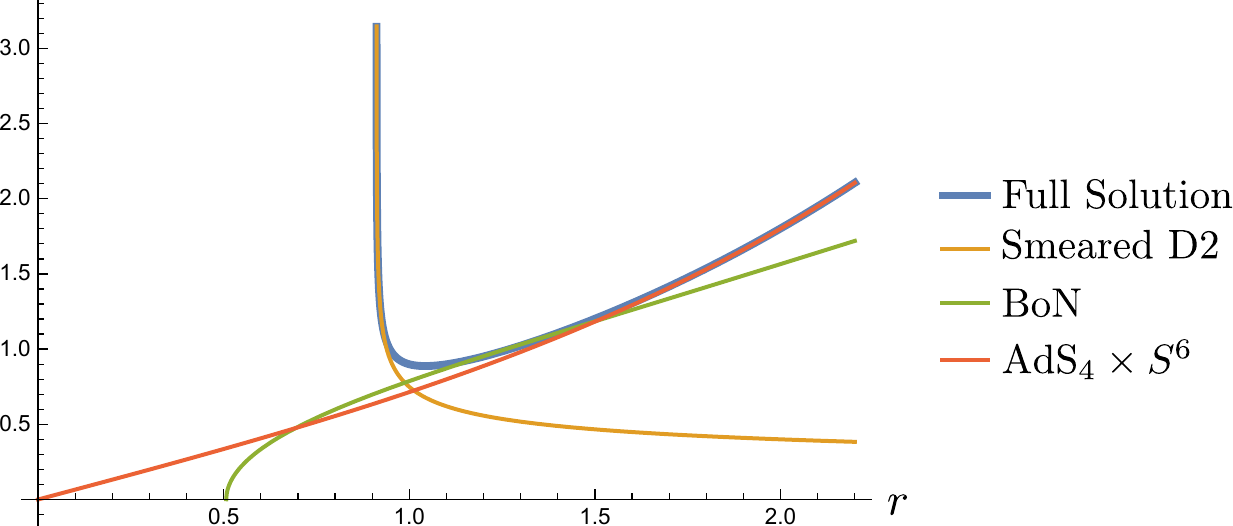}
	\caption{\it The profile of the factor $e^{A+U+\frac{\phi}{4}}$ in front of dS$_3$. Plot taken from \cite{bubbles}.}
	\label{fig:Regimes_U}
\end{figure}

\begin{figure}[!http]
	\centering
	\includegraphics[width=0.75\textwidth]{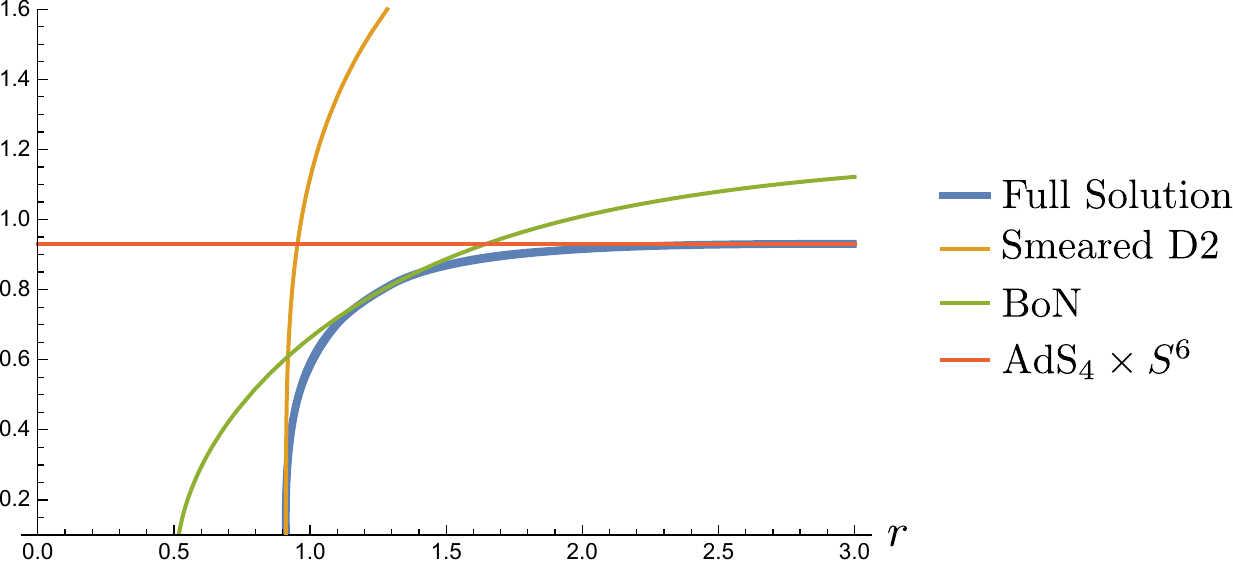}
	\caption{\it  The radial profile of the factor $e^{U+\frac{\phi}{4}}$ in front of $S^6$. Plot taken from \cite{bubbles}.
	}
	\label{fig:Regimes_W}
\end{figure}

\begin{figure}[!http]
	\centering
	\includegraphics[width=0.75\textwidth]{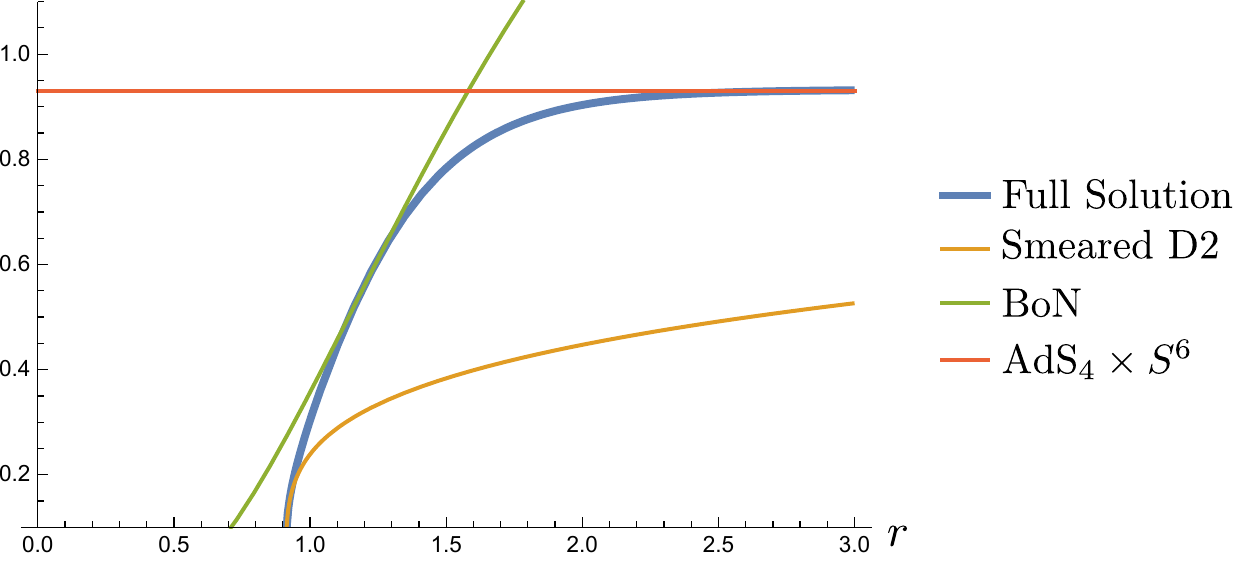}
	\caption{\it  The radial profile of the 10d dilaton $e^{\phi}$. Plot taken from \cite{bubbles}.
	}
	\label{fig:Regimes_phi}
\end{figure}

Finally, in figures \ref{fig:Regimes_U}, \ref{fig:Regimes_W} and \ref{fig:Regimes_phi} we plot the numerical solution (blue curves) together with the three different regimes (we use respectively red, yellow and green curves to plot the AdS$_4$ vacuum, the D2 source and the intermediate bubble regime).

\newpage

\section*{Acknowledgements}

I would like to acknowledge my collaborators Pieter Bomans, Davide Cassani, Giuseppe Dibitetto and Marjorie Schillo for the collaboration in the works \cite{Dibitetto:2020csn} and \cite{bubbles} whose results are reviewed in this article. I especially thank Giuseppe Dibitetto for our long-time collaboration and for his valuable mentorship. This work is supported by the Israel Science Foundation (grant No. 741/20) and by the German Research Foundation through a German-Israeli Project Cooperation (DIP) grant "Holography and the Swampland".

\bibliographystyle{utphys}
  \bibliography{references}
\end{document}